# Reimagine Procrastination: Music Preference and Health Habits as Factors on Self-Perceived Procrastination of Young People

Shu Wei (ucfnswe@ucl.ac.uk)
MRes Spatial Data Science and Visualisation
The Bartlett Centre for Advanced Spatial Analysis

**ABSTRACT**
As the buzzword phenomenon, procrastination holds a continued need for a comprehensive examination of its nature and the associated factors. The presented study explores the potential relationship between music taste, living style and the youngsters' procrastination through quantitative modelling. To handle the big set of survey statistics and the uncertainty caused by the data missingness, the combined methods of factor analysis, multiple imputation (MI) and ordered logit regression are employed. The result reveals that the music preference for Hip-hop, AlternativeR and Opera have a significant effect on procrastination. Concerning the living habits, the eating habit and local authority (city/rural) also yield strong connection to the self-perceived procrastination. Implications for this procrastination research is discussed.

**INTRODUCTION**
Procrastination is highly embedded in young people's life with a continuous growing prevalence (Tice & Baumeister, 1997; Steel, 2007; Sirois & Pychyl, 2016). Defined by Jaffe (2013) as the "voluntary delay of some important task that we intend to do, despite knowing that we'll suffer as a result", it mostly appears as a troubling phenomenon, and more than 90% procrastinators yield their intentions to overcome it (O'Brien, 2002). Even though a great number of studies have shed light on understanding procrastination and its empirical solutions (Perry, 2014; Sirois & Pychyl, 2016), there is a continued need for a comprehensive examination of its nature, especially the factors associated with procrastination. The current study focuses on the music preference and health habits as two main parts of the factors, exploring what is the potential relationship between music taste, lifestyle and the self-perceived procrastination tendency of young people by quantitative modelling.

**Music Preference and Working Style**
Musical tastes sometimes serve as a great window into how people think and behave (Greenberg et al., 2015). The online personality test website 16 Personality (2017) presented its report analysing the link between music preference and characteristics based on the music taste tests of over 4000 respondents. Combined with their reported lifestyle, it suggested that people who like to listen to Classical and Alternative Rock music have a significantly higher chance (>74%) to be more assertive in action compared to other music choices. The online data-based insights supported Blacking's (1995) idea that we decide who we are partly by what music we listen to. People use their own empathy-based judgment to react to different musical contents emotionally, thus musical preferences reflect explicit characteristics and personal styles (Greenberg et al., 2015). However, it still worth noticing that the individual differences account for stronger relation to personality than music taste, and the power of music as a vessel for ideas is better realized when people partake their preference to others consciously (North, 2010; 16 Personality, 2017).

**Health and Procrastination**
A growing body of research revealed both physical health and mental wellbeing are linked tightly to procrastination (Sirois & Pychyl, 2016).

One of the early investigations rated university students to an established procrastination scale, and then recorded their academic performance and general health condition throughout the term (Tice & Baumeister, 1997). Even though the procrastinators initially reported to be beneficial from their procrastination for less pressure, in the end of the term, the lower grades and higher cumulative amounts of stress and illness were proved to associated with high-level procrastination. However, Perry (2014) pointed out procrastination could also be valuable in long-term. Since a great number of procrastinators often fantasize about doing things perfectly, they made the intentional delay to waiting for more information come along to achieve a better result. In this sense, procrastination may result in a better life quality and higher self-satisfaction in terms of psychological health (Bernstein, 1998).

## DATA

The chosen dataset is the Young People Survey (N=1009, mean age=23) conducted by the Comenius University in 2013. The survey consists of 149 items including a set of self-measured statements and personal demographics information.

To explore the potential association of music preference and health habits with procrastination, the raw data is subset and re-sorted to 31 items. The likelihood scale of self-rated statement "I try to do tasks as soon as possible and not leave them until last minute" is reversed and treated as the dependent variable representing the self-perceived procrastination (1 means not procrastinated at all, 5 means strongly procrastinated). The remaining items include the preference of 16 music categories, 5 health-related habits and 9 demographics data.

## METHOD

### Variable Selection: Factor Analysis and Chi-Square Test

Considering the possible multicollinearity effect that one predictor variable may be linearly correlated with others among the above variables, a factor analysis is conducted to screen for the redundant data for the more effective multiple regression model (Farrar & Glauber,1967).

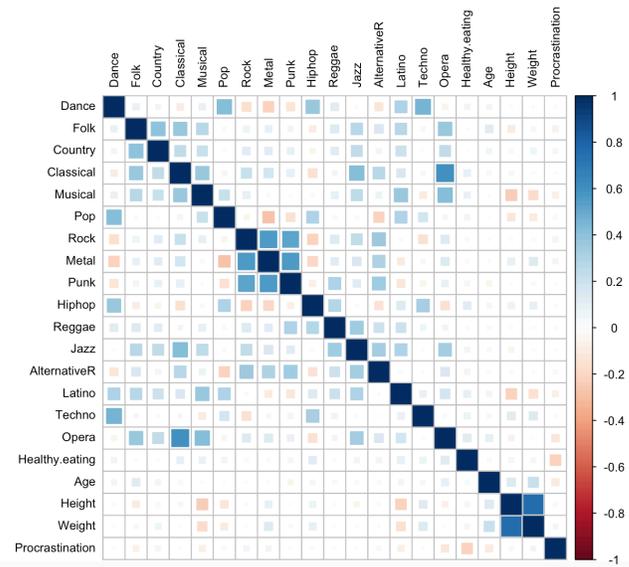

Figure 1. correlation plot of non-categorical variables

All the non-categorical data are input for the Spearman rank-order correlation test, which is robust for detecting the relationship between both continuous and ordinal variables. The coefficient visualisation (figure 1) shows that the height and weight are highly correlated (r = 0.76), therefore, the concept of body mass index (BMI= weight (kg)/height^2(m)) is introduced to combine both value, demonstrating a rough clue about the individual obesity and physical health (Taylor, 2010). While the preference for the music group of Rock, Metal and Punk as well as Opera and Classical exhibit a moderate correlation (r ≈ 0.5), these genres are still kept separated since music preference is part of the main fields of interest.

To select the complementary demographics data which is in categorical form, a Chi-square test is conducted to test the null hypothesis that there is no significant difference of how different groups perceive their procrastination ($\partial$=0.05).

The above manipulations leave the predictors variables to the overall size of 26, where three binary items of "gender", "left or right hand" and "only child" are also deleted (table 1).

| Sub-part | Music Preference | Health Habits & life style | Demographics |
|---|---|---|---|
| Variable | Ordinal:<br>Dance<br>Folk<br>Country<br>Musical<br>Pop<br>…<br>Reggae<br>Jazz<br>AlternativeR<br>Latino<br>Opera | Categorical:<br>Smoking habit<br>Drinking habit<br>Internet habit<br>Punctuality habit<br><br>Ordinal:<br>Self-rated eating health level | Numerical:<br>Age<br>BMI (weight/height^2)<br><br>Categorical:<br>Education level<br><br>Local authority (village or town)<br><br>House type (block of flats or house/bungalow) |
| Number of variable | n=16 | n=5 | n=5 |

Table 1. predictor variables overview

**Multiple Imputation for Missing Data**

Based on the nature of the survey statistics, one big challenge is to handle the big amount of missing data ($n_{NA}$= 149) existing among all the predictor items. While the common practise such as step-wise selection gives a solution by simply deleting the missing data list-wise, it suffers from the loss of important information and leads to bias in quantitative modelling (Pampaka, Hutcheson & Williams, 2014). Therefore, a more sophisticated multiple imputation (MI) method is adopted here.

The key concept of MI is to appropriately estimate the uncertainty of missing data by computing a set of plausible values, then perform the statistical analysis on each individual imputed model, and combine the multiple outputs to a better-adjusted result (Little & Rubin, 1987). By employing this general idea, the presented analysis is conducted by the following four steps:
(1) Examine the missing data mechanism to detect any special missing data pattern
(2) Compute the imputed dataset iteratively (5 times) using the appropriate model (EBM model here) to get multiple versions of completed data
(3) Perform the ordinal logistic regression (discussed in next section) separately on each dataset
(4) Combine all the statistics results by taking the average value of the single figure. Additionally, to incorporate the variability of repeated imputation, the standard error here is calculated by getting the mean of squared standard errors and variance of the N set of estimates, and then combining the two values with an adjustment term (1+1/N) (a) (Durrant, 2009; Pampaka, Hutcheson & Williams, 2014).

$$SE = \sqrt{\frac{1}{N}\sum_i S_i^2 + (1 + \frac{1}{N})(\frac{1}{N-1})\sum_i (c_i - \overline{c})^2} \quad (a)$$

The above procedures are conducted within the software environment of R. After identifying the dataset has a multivariate normal distribution, and the omitted data is missing completely at random (MCAR), the MI is performed using the Amelia II package (Honaker, King & Blackwell, 2011). The multiple imputation algorithm in Amelia II is based on the EMB model which combines Expectation-Maximization (EM) algorithm in a bootstrap approach and Bayesian Hierarchical classification model for the missing value estimation (Antunes, Cardoso & Pinto, 2010).

**Ordered Logit Regression of Imputed Data**

Considering the ordinal feature of the response variable self-perceived procrastination with a scale from 1 to 5, the ordered logit regression model (also proportional odds model) is chosen to estimate the probability of the procrastination response falling into a certain level given a set of predictors (PSECS, 2017). The ordered logit model is based on the proportional odds assumption and uses the cumulative logit to describe the log-odds of two cumulative probabilities (b). Therefore, the logarithms of the odds of response of certain procrastination level are shown in table 2.

$$\log\left(\frac{P(Y \leq j)}{P(Y > j)}\right) = \log\left(\frac{P(Y \leq j)}{1 - P(Y \leq j)}\right) = \log\left(\frac{p_1 + \cdots + p_j}{p_{j+1} + \cdots + p_j}\right) \quad (b)$$

(where Y=1,2,…j, and the associated probabilities are {$p_1$, $p_2$,…, $p_j$})

| Response of procrastination level | Formula | Simple sequence |
|---|---|---|
| 1 | $\log\left(\frac{p_1}{p_2+p_3+p_4+p_5}\right)$ | 1 |
| 1 or 2 | $\log(\frac{p_1+p_2}{p_3+p_4+p_5})$ | 2 |
| 1, 2, or 3 | $\log(\frac{p_1+p_2+p_3}{p_4+p_5})$ | 3 |
| 1, 2, 3 or 4 | $\log(\frac{p_1+p_2+p_3+p_4}{p_5})$ | 4 |

Table 2

The above equations give out the log-odds of the ordered response falling into or below category *j* versus falling above it. The model can be further simplified by requiring the coefficient (β) of each X variable to be identical across the N-1 logit equations. Then the probability logit and probability of simple sequence N-1 can be represented as (c) and (d) respectively, where α is the intercept and i is the number of X variable.

$$\log[P(N-1)] = \alpha_{N-1} + \beta_1 X_1 + \cdots + \beta_i X_i \quad (c)$$

$$P(N-1) = \frac{e^{(\alpha_{N-1}+\beta_1 X_1+\cdots+\beta_i X_i)}}{1+e^{(\alpha_{N-1}+\beta_1 X_1+\cdots+\beta_i X_i)}} \quad (d)$$

The ordered logit regression is run on the multiple imputed data within R, utilizing the Zelig package (Venables & Ripley, 2011) combined with Amelia II (Honaker, King & Blackwell, 2011). A self-defined function is also created to loop through the multiple outputs and calculate the combined estimates statistics as a final result.

**RESULT**

Given the final sample size n=1005 in each individual dataset. The response of the self-rated procrastination is shown as figure2.

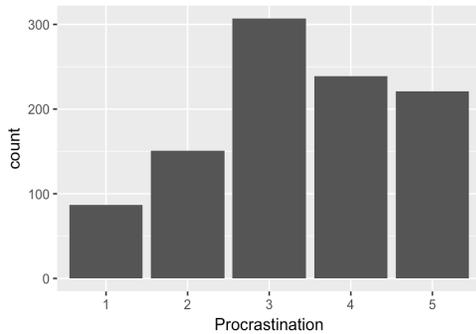

Figure 2

By dropping out all the insignificant predictor variables based on the combined result from 5 imputed data set (∂=0.05), the music preference for "Hip-hop", "Alternative R" and "opera", the self-rating for "healthy eating habit", the specific type of "smoking ('tried smoking')", "alcohol habits ('drink a lot')" and the "local authority ('city')" yield significant association with the perceived procrastination in the final model (table3).

| Coefficients: | Value | Std. Error | t value | p value |
|---|---|---|---|---|
| Hiphop | -0.064 | 0.044 | -1.451 | 0.046 |
| AlternativeR | 0.105 | 0.044 | 2.396 | 0.017 |
| Opera | -0.121 | 0.049 | -2.479 | 0.013 |
| Healthy.eating | -0.123 | 0.064 | -1.921 | 0.045 |
| Smoking ("tried smoking") | 0.337 | 0.155 | 2.179 | 0.029 |
| Alcohol ("drink a lot") | 0.360 | 0.215 | 1.676 | 0.024 |
| LocalAuthority ("city") | -0.148 | 0.155 | -0.953 | 0.034 |

Table 3

| Intercepts: | Value | Std. Error | t value |
|---|---|---|---|
| 1\|2 | -6.4771 | 1.3564 | -4.7753 |
| 2\|3 | -5.1846 | 1.3509 | -3.8378 |
| 3\|4 | -3.6693 | 1.3467 | -2.7246 |
| 4\|5 | -2.4174 | 1.3442 | -1.7984 |

Table 4

The result indicates that in terms of music preference, for one unit increase in the Likert scare for Hiphop, AlternativeR and Opera, we expect about the value changes of -0.064, 0.105 and -0.121 respectively in the expected value of procrastination in the log-odds scale, given the remaining variables in the model are held constant. Also, when the individual selects the choice of 'tried smoking', 'drink a lot', 'city' in terms of their smoking, drinking habit and local authority, we expect about the value changes of 0.337, 0.360 and -0.148 respectively in the expected value of procrastination in the log-odds scale (IDRE, 2017). Based on the above estimates, we can get the prediction equations as below:

($y = -0.064 X_{hipop} + 0.105 X_{AlternativeR} - 0.121 X_{opera} - 0.123 X_{healthyEating} + 0.337 X_{tried\ smoking} + 0.36 X_{drink\ a\ lot} - 0.148 X_{city}$)

$$p(1) = \frac{1}{1 + e^{-(-6.48-y)}}$$

$$p(1|2) = \frac{1}{1+e^{-(-5.1846-y)}} \rightarrow p(2) = p(1|2) - p(1)$$

...

$$p(1|2|3|4|5) = 1 \rightarrow p(5) = 1 - p(1|2|3|4)$$

## DISCUSSION AND CONCLUSION

The presented study utilizes the proportional-odds cumulative logit regression to project the ordinal response of young people's perceived procrastination, given the predictors including music preference, health habits and local authority. Even though the statistical result is less intuitive to interpret, and the coefficient does not give out a direct relationship between the X and dependent variable, the proportional-odds assumption is considered robust for analysing attitude responses by taking mixed types of explanatory factors (Larasati, DeYong & Slevitch, 2011). The McFadden's Pseudo-$R^2$ [1] is calculated to evaluate the goodness of the model, and the $R_M^2$ value of 0.31 suggests the model has a moderate fitness (Agresti, 1996). Analysis is computed iteratively on the multiple imputed datasets, which reduces the bias due to the big data missingness of Young People Survey, while also lowering the risk of the uncertainty of the estimation for the subjective self-rating response (Pampaka, Hutcheson & Williams, 2014). Even though the repeated imputation with EMB algorithm is computationally expensive and increases the complexity of generating the final result with different sets of estimates, it is still highly encouraged to implement utilising the easy-to-access software (Pampaka et al., 2013).

There are 7 out of 26 variables being employed in the final model, considered to be significantly associated with the perceived procrastination. The music preference for Hip-hop, AlternativeR and Opera are identified as effective factors on procrastination. Also, people who perceive a same level of eating health are more likely to fall into a specific level of procrastination. Additionally, the local authority of the young people yields a statistically strong connection to the procrastination, especially for those who live in city. Interestingly, the items such as punctuality level, online time and gender, which were considered to correlated to procrastination proved to be statistically insignificant (Steel, 2007). However, it worth noticing that the chosen data sample may not be representative enough to generalise to the broad young people group, considering all the participants of the survey are Slovak nationals. Furthermore, the individual difference could introduce great uncertainty and potential error for the quantitative model, especially when both the perceived procrastination and music preference are subjective (North, 2010).

Nevertheless, the empirical research successfully provides insights for the buzzword-phenomenon procrastination through the associated factors of life style and music preference, linking human behaviours towards different issues based on their cognitive nature. Further exploration could be undertaken by utilizing the longitudinal experiments to examine whether the adoption of healthy eating, listening to specific music (ie. Opera) and certain living environment can effectively reduce the procrastination tendency, serving as the remedy for the epidemic youngster's disease. (2012 words)

---

[1] $R_M^2 = 1 - \frac{\ln \hat{L}_{full}}{\ln \hat{L}_{intercept}}$

# APPENDIX
# Technical Implementation in R

Missing data imputation

```
impDatasets <- amelia(varibleSubset1, m=5, frontend =FALSE,p2s=2,
         noms=c("SmokingCoded","AlcoholCoded","PunctualityCoded",
          "EducationCoded", "VillageorTownCoded","HouseCoded","OnlineCoded"),
         ords=c("Dance","Folk","Country","Classical","Musical","Pop","Rock","Metal",
                "Punk","Hiphop","Reggae","Jazz","AlternativeR","Latino","Techno",
          "Opera","Healthy.eating","Age"),
         idvars=c("ProcrastinationCoded","Procrastination"))

write.amelia(obj=impDatasets, file.stem = "outdata")
```

Ordinal logistic regression

```
install.packages("zeligverse")
library(zeligverse)
library(Zelig)

logicModel<- zelig(as.factor(ProcrastinationCoded)~Dance+Folk+Country+Classical+
                Musical+Pop+Rock+Metal+Punk+Hiphop+Reggae+
                Jazz+AlternativeR+Latino+Techno+Opera+Healthy.eating+
                Age+BMI+SmokingCoded+AlcoholCoded+OnlineCoded+PunctualityCoded+
                EducationCoded+VillageorTownCoded+HouseCoded,
                model = "ologit", data =impDatasets$imputations)
summary(logicModel,subset=1:5)
```

Estimates combination of multiple imputation

```
a<-NULL
b<-NULL
for(i in 1: impDatasets$m) {
  log.out <- zelig(as.factor(ProcrastinationCoded)~Hiphop+Healthy.eating+Age+Height+
                AlcoholCoded+PunctualityCoded+VillageorTownCoded,
                model = "ologit", data =impDatasets$imputations[[i]])

  a <- log.out$get_coef()
  b <- rbind(b, coefficients(summary(log.out))[,2])
}
  combined.results <- mi.meld(q = a, se = b)
```